\documentclass[preprintnumbers,amsmath,amssymb]{revtex4} 

\usepackage{graphicx}
\usepackage{dcolumn}
\usepackage{bm}
\usepackage{fancyhdr, lastpage} 

\begin{document}

\title{A simple eddy viscosity formulation for turbulent boundary layers near smooth walls}

\author{Rafik Absi} 
  \email{r.absi@ebi-edu.com}
\affiliation{
EBI, Inst. Polytech. St-Louis, 32 Boulevard du Port, 95094 Cergy-Pontoise Cedex, France.\\ 
At the time of submission, visiting researcher at St. Anthony Falls Laboratory, University of Minnesota, Minneapolis, USA. 
}%

\begin{abstract}
The aim of this study is to improve the prediction of near-wall mean streamwise velocity profile $U^+$ by using a simple method. The $U^+$ profile is obtained by solving the momentum equation which is written as an ordinary differential equation. An eddy viscosity formulation based on a near-wall turbulent kinetic energy $k^+$ function (R. Absi, Analytical solutions for the modeled $k$-equation, ASME J. Appl. Mech. \textbf{75}, 044501, 2008) and the van Driest mixing length equation (E.R. van Driest, On turbulent flow near a wall, J. Aero. Sci. \textbf{23}, 1007, 1956) is used. 
The parameters obtained from the $k^+$ profiles are used for the computation of $U^+$ (variables with the superscript of $+$ are those nondimensionalized by the wall friction velocity $u_\tau$ and the kinematic viscosity $\nu$). Comparisons with DNS data of fully-developed turbulent channel flows for $109 < Re_{\tau} < 2003$ show good agreement (where $Re_{\tau}$ denotes the friction Reynolds number defined by $u_\tau$, $\nu$ and the channel half-width $\delta$). 
\end{abstract}

\maketitle

\begin{center}
\textbf{NOMENCLATURE} 
\end{center}
$A_k^+$, $A_l^+$, $B$, $C$, $C_{\nu}$ = coefficients \\ 
$k$ = turbulent kinetic energy \\ 
$l_m$ = mixing length \\ 
$P$ = pressure \\ 
$Re_{\tau}$ = friction Reynolds number \\ 
$x$, $y$ = coordinates in respectively the streamwise and wall normal directions \\ 
$U$, $V$ = mean velocity components respectively in the $x$ and $y$ directions \\ 
$u_\tau$ = wall friction velocity \\ 
$\delta$ = channel half-width \\ 
$\kappa$ = K\'arm\'an constant ($\approx 0.4$) \\ 
$\nu$ = kinematic viscosity \\ 
$\nu_t$ = eddy viscosity \\ 
$\rho$ = density \\ 
$\tau$ = shear stress \\ 
All variables with the superscript of $+$ are those nondimensionalized by $u_\tau$ and $\nu$ \\ 

\section{Introduction}

Turbulent flows are significantly affected by the presence of walls 
\cite{Hinze}. 
Successful predictions of turbulence models used for wall-bounded turbulent flows depend on accurate description of the flow in the near-wall region. 
Numerous experiments 
of fully-developed turbulent channel flows, show that the near-wall region can be subdivided into three layers. 
A viscous sublayer (for a distance from the wall $y^+ < 5$), where the mean velocity $U^+$ can be approximated by $U^+ = y^+$ 
and the turbulent kinetic energy $k^+$ by a quadratic variation $k^+ \approx y^{+ 2}$ \cite{Hanjalic}. 
A fully-turbulent layer or outer layer (for $y^+ > 30$ until an upper limit), where $U^+$ can be correctly approximated by the logarithmic profile \cite{Tennekes} and $k^+$ by an exponential decaying function \cite{AbsiJAM}. 
Between these two layers, a buffer layer, where $k^+$ can be accurately predicted by an analytical function \cite{AbsiJAM}. 

The aim of this Note is to improve the prediction of $U^+$ 
by using a simple and accurate method. 
The $U^+$ profile will be obtained from the resolution of the momentum equation. 
An eddy viscosity formulation based on a near-wall turbulent kinetic energy $k^+$ function \cite{AbsiJAM}, which was validated by DNS data for $109 < Re_{\tau} < 642$ 
for $y^+ < 20$, and the van Driest mixing length equation will be used. 
The values of $U^+$ and $k^+$ at an upper limit of the buffer layer could be used as boundary conditions for a turbulence closure model 
applied in the outer layer.\\

The test case is the fully developed plane channel flow which is considered to be the simplest and most idealized boundary layer flow. 
Reynolds number effects on wall turbulence have been investigated by many experimental and computational studies. 
A review of turbulence closure models for wall-bounded shear flows was presented in Patel \textit{et al.} (1985) \cite{Patel}, and experiments in the range of $190 < Re_{\tau} < 1900$  
was performed by Wei and Willmarth (1989) \cite{Wei} to investigate the effects of the Reynolds number very near the wall. 
There are several DNS studies of plane channel flows which have allowed to improve the knowledge of the boundary layer dynamics. The DNS was performed at $Re_{\tau} = 180$ by Kim \textit{et al.} (1987) \cite{Kim}, up to $Re_{\tau} = 590$ by Moser \textit{et al.} (1999) \cite{Moser}, up to $Re_{\tau} = 642$ by Iwamoto \textit{et al.} \cite{Iwamoto} (2002), up to $Re_{\tau} = 950$ by del \'Alamo \textit{et al.} (2004) \cite{delAlamo}, and recently at $Re_{\tau} = 2003$ by Hoyas and Jim\'enez (2006) \cite{Hoyas}.


\section{Model equations}

\begin{figure}
\includegraphics[width=10cm,height=6cm]{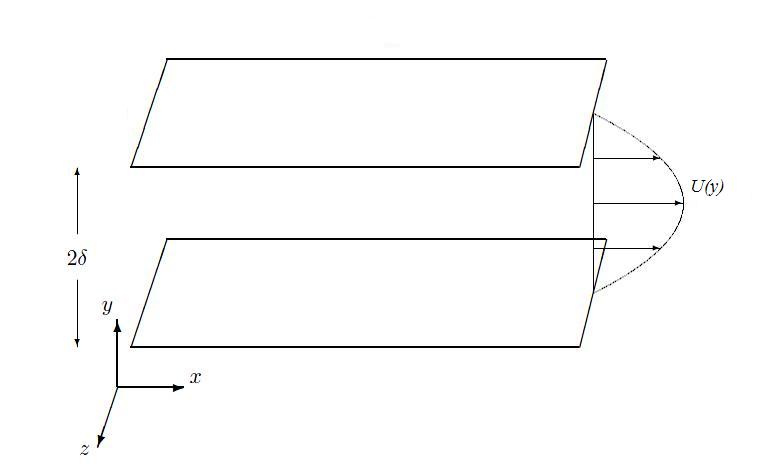}  
\caption{\label{fig:Sketch} Sketch of the flow geometry for plane channel flow. 
}
\end{figure}

We consider a steady uniform fully developed plane channel flow (i.e. the flow between two infinitely large plates, fig.~\ref{fig:Sketch}), where $x$ and $y$ are respectively the coordinates in the streamwise and wall normal directions and the corresponding mean velocity components are respectively $U$ and $V$. The channel half width is $\delta$ (can represent the boundary layer thickness), and the flow is driven by a pressure gradient in the streamwise direction.

\subsection{Momentum equation} 

\begin{figure}
\includegraphics[width=8cm,height=12cm]{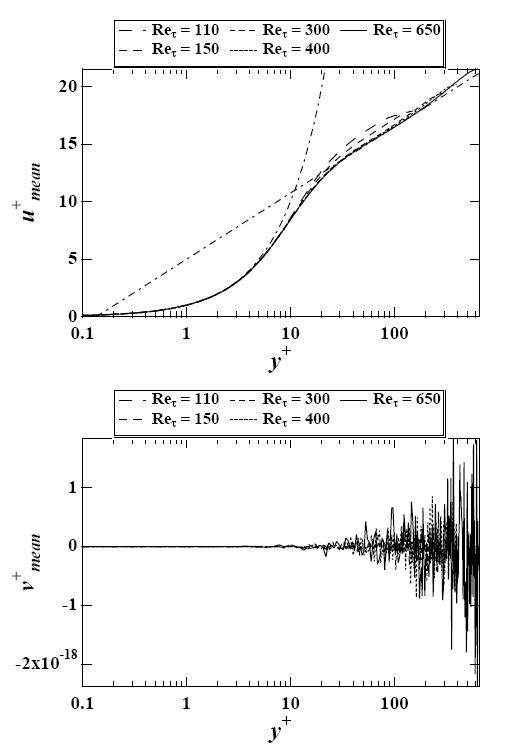}  
\caption{\label{fig:UpVp} DNS data \cite{Iwamoto} of mean velocity profiles for $109 < Re_{\tau} < 642$. Bottom figure, $v_{mean}^+(y^+)=V^+(y^+)$; Top figure, $u_{mean}^+(y^+)=U^+(y^+)$, dash-dotted lines, $U^+ = y^+$ and $U^+ = 2.5 \:  ln(y^+)+5.0$ (figure from \cite{Iwamoto2}). 
}
\end{figure}

DNS data (Fig.~\ref{fig:UpVp}) \cite{Iwamoto} for $109 < Re_{\tau} < 642$ show that $V \approx 0$ for $y^+ < 20$. 
By taking $V = 0$, 
the streamwise momentum equation becomes 
\begin{eqnarray}
(1 / \rho) \partial_x P = \partial_y \left( (\nu + \nu_t) \partial_y U \right)
\label{NS3}
\end{eqnarray}
where $\nu_t$ is the eddy viscosity, $P$ the pressure and $\rho$ the density. With the shear stress $\tau$, we write Eq. (\ref{NS3}) as 
\begin{eqnarray}
\partial_x P = \partial_y \tau 
\label{NS4}
\end{eqnarray}
where $\tau = \rho \: (\nu + \nu_t) \partial_y U$. 
For a constant $\partial_x P$, by integrating Eq. (\ref{NS4}) between $\tau(y=0) = \tau_w$ and $\tau(y=\delta) = 0$, we obtain 
$\tau_w = - \delta \: \partial_x P$ and therefore $u_{\tau} =   \sqrt{(\delta/\rho) (- \partial_x P)}$.

By integrating Eq. (\ref{NS3}) between $y=0$ and $y=\delta$ we obtain 
\begin{eqnarray}
\frac{d U}{d y}  = \frac{u_{\tau}^2}{\nu + \nu_t} \: \left(1 - \frac{y}{\delta}\right)
\label{NS5}
\end{eqnarray}

Or in wall unit 
\begin{eqnarray}
\displaystyle \frac{d U^+}{d y^+} = \frac{1} {1 + \nu_t^+} \: \left( 1 - \frac{y^+} {Re_{\tau}}  \right) 
\label{NSF}
\end{eqnarray}
where $U^+ = U / u_{\tau}$, $y^+ = y \: u_{\tau} / \nu$ and $\nu_t^+ = \nu_t / \nu$. 
The resolution of the ordinary differential equation (\ref{NSF}) needs the dimensionless eddy viscosity $\nu_t^+$.

\subsection{A near-wall eddy viscosity formulation} 

The eddy viscosity is given by 
\begin{eqnarray} 
\displaystyle \nu_t = C_{\nu} \sqrt{k} \: l_m
\label{nutu}
\end{eqnarray}
where $l_m$ is the mixing length and $C_{\nu}$ a coefficient. 

On the one hand, the mixing length is given by the van Driest equation 
\begin{eqnarray}
\displaystyle l_m = \kappa y \left( 1 - e^{\displaystyle - y^+ / A_l^+}  \right) 
\label{lmVD}
\end{eqnarray}
where $\kappa$ is the K\'arm\'an constant ($\approx 0.4$) and $A_l^+ = 26$. 
We write $\nu_t^+$ from equations (\ref{nutu}) and (\ref{lmVD}) as 
\begin{eqnarray}
\displaystyle \nu_t^+ = \frac{\nu_t}{\nu} 
= C_{\nu} \sqrt{k^+} \kappa \frac{y  u_{\tau}}{\nu} \left( 1 - e^{\displaystyle - y^+ / A_l^+}\right) 
= C_{\nu} \sqrt{k^+} \: l_m^+
\label{nutplus}
\end{eqnarray}
where $k^+ = k / u_{\tau}^2$ and $l_m^+ = \kappa y^+ \left( 1 - e^{\displaystyle - y^+ / A_l^+}\right)$.


\begin{figure}
\includegraphics[width=12cm,height=10cm]{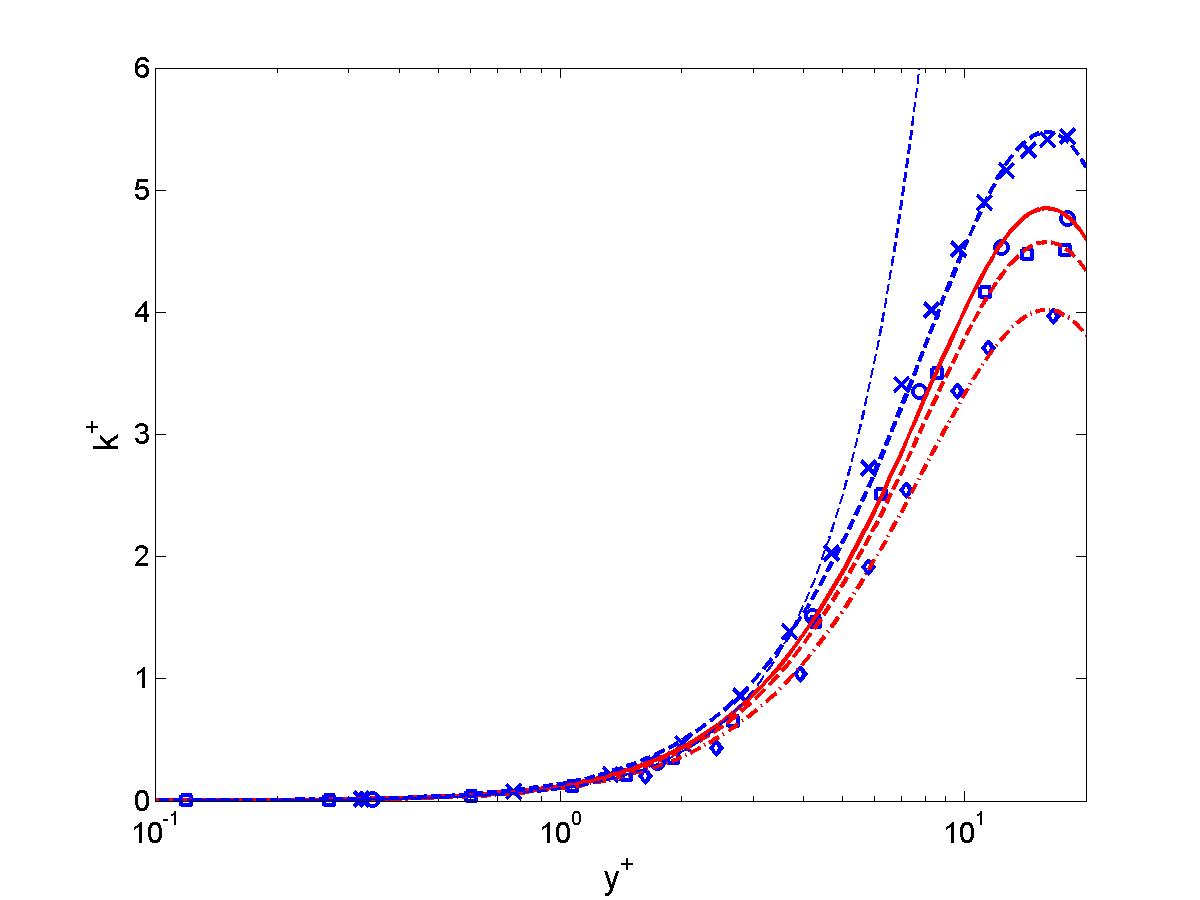}  
\caption{\label{fig:kpyp} Turbulent kinetic energy $k^+(y^+)$ for different Reynolds numbers (for $y^+ < 20$). 
Symbols, DNS data. 
$Re_{\tau} = 150$, diamonds \cite{Iwamoto}, dash-dotted line Eq. (\ref{kpluswall}) with $A_k^+=8$ and $B=0.116$; 
$Re_{\tau} = 395$, squares \cite{Iwamoto}, dashed line Eq. (\ref{kpluswall}) with $A_k^+=8$ and $B=0.132$; 
$Re_{\tau} = 642$, circles \cite{Iwamoto}, solid line Eq. (\ref{kpluswall}) with $A_k^+=8$ and $B=0.14$; 
$Re_{\tau} = 2003$, $\times$ \cite{Hoyas}, dashed line Eq. (\ref{kpluswall}) with $A_k^+=8$ and $B=0.158$; 
Thin dashed line, $k^+=0.1 y^{+ 2}$. 
}
\end{figure}


On the other hand, from the modeled $k$-equation, we developed a function for $k^+$ for $y^+ < 20$ \cite{AbsiJAM}. 
For steady channel flows, we write the $k$-equation as 
$\displaystyle \partial_y \left( \nu_t \partial_y k \right) = - \left(G + \partial_y \left( \nu \: \partial_y k \right) - \epsilon \right)$, where $G$ and $\epsilon$ are respectively the energy production and dissipation. With an approximation for the right-hand side as $\left(G + d_y \left( \nu \: d_y k \right) - \epsilon \right) \approx 1 / y^2$ and by integrating, we obtained \cite{AbsiJAM} 
\begin{eqnarray}
\displaystyle k^+ = B \: y^{+ 2 C} \: e^{\displaystyle( - y^+ / A_k^+)}  
\label{kplus}
\end{eqnarray} 
Where $A_k^+$, $B$ and $C$ are coefficients. 
Examination of Eq. (\ref{kplus}) by DNS data of channel flows 
shows that for $y^+ \leq 20$, $C = 1$, $A_k^+ = 8$ and $B$ is $Re_{\tau}$-dependent \cite{AbsiJAM}. 
We write therefore $k^+$ for $y^+ \leq 20$ as  
\begin{eqnarray}
\displaystyle k^+ = B \; y^{+ 2} \; e^{\displaystyle( - y^+ / A_k^+)} 
\label{kpluswall}
\end{eqnarray} 
Table \ref{tab:table1} gives values of $B(Re_{\tau})$ obtained from Eq. (\ref{kpluswall}) and DNS data \cite{Iwamoto}, \cite{Hoyas}, \cite{AbsiJAM}. 

We propose the following function Eq. (10) for the coefficient B 
\begin{eqnarray}
B(Re_{\tau}) = C_{B1} ln(Re_{\tau}) + C_{B2} 
\label{B}
\end{eqnarray} 

where $C_{B1}$ and $C_{B2}$ are constants. The calibration (Fig. 4) gives $C_{B1} = 0.0164$ and $C_{B2} = 0.0334$.

\begin{table} 
\caption{\label{tab:table1}Values of coefficient $B(Re_{\tau})$ obtained from Eq. (\ref{kpluswall}) and DNS data.} 
\begin{ruledtabular}
\begin{tabular}{c |c c c c c c}
$Re_{\tau}$ &109 &150 &298 &395 & 642 & 2003 \\
\hline
$B$ &0.11 &0.116 &0.127 &0.132 & 0.14 & 0.158\\ 
\end{tabular}
\end{ruledtabular}
\end{table}

\begin{figure}
\includegraphics[width=12cm,height=10cm]{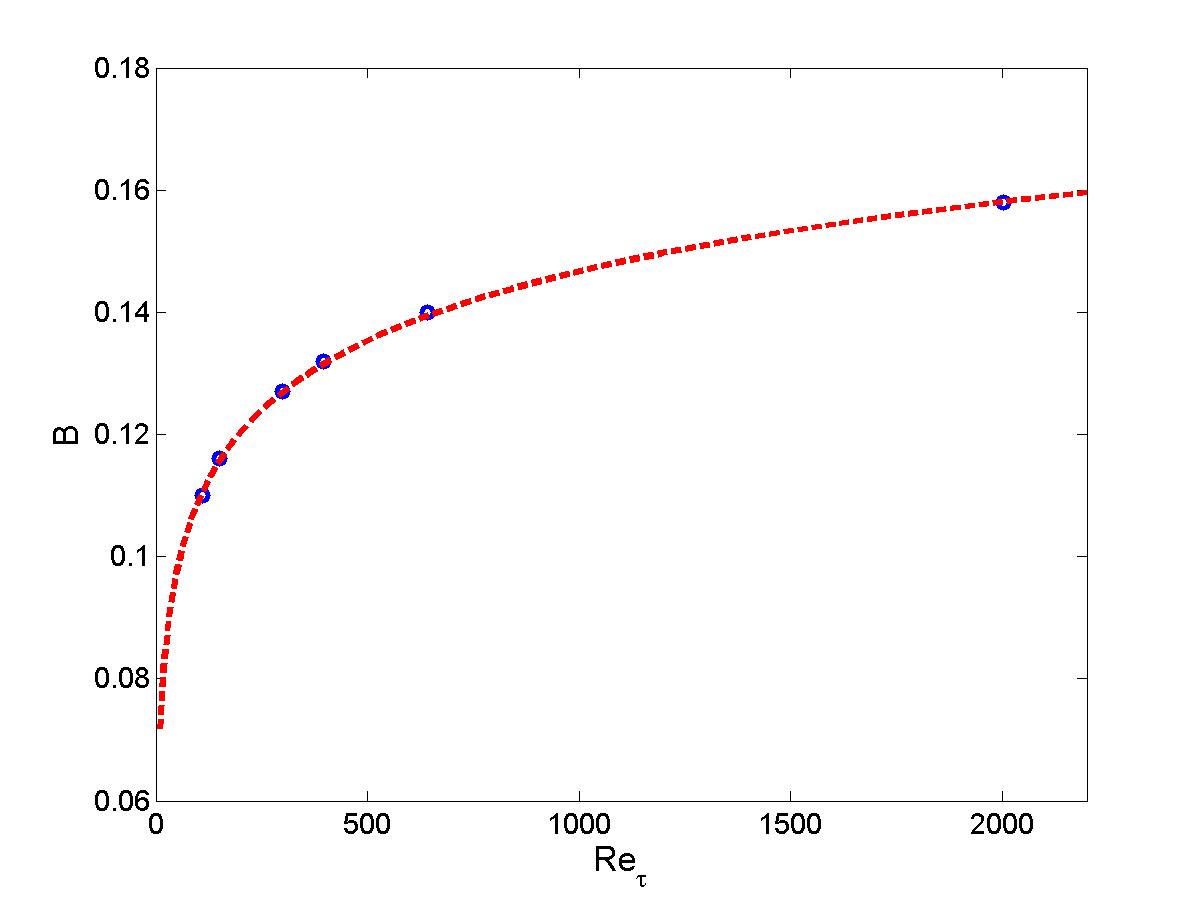}  
\caption{\label{fig:Usol} Dependency of the coefficient B on the Reynolds number $Re_{\tau}$. o, values obtained from DNS data; Curve, proposed function (10). 
}
\end{figure}

We noticed that the series expansion of the exponential in Eq. (\ref{kpluswall}) at the first order gives $k^+=B y^{+ 2}-(B / A_k^+) y^{+ 3}$. This equation is similar to the approximation deduced from the continuity equation and the no-slip condition
\cite{Hanjalic} (page 608). However, the quadratic variation of $k$ (first term in the right-hand side) is valid only in the immediate vicinity of the wall ($y^+ < 5$). Eq. (\ref{kpluswall}) is therefore a more general and more accurate solution (Fig.~\ref{fig:kpyp}).\\

With Eq. (\ref{kpluswall}), we write the dimensionless eddy viscosity as 
\begin{eqnarray}
\displaystyle \nu_t^+ 
= C_{\nu}  \: \kappa  \: B^{0.5} \: y^{+ 2}  \: e^{\displaystyle - y^+ / (2 A_k^+)} \left( 1 - e^{\displaystyle - y^+ / A_l^+}\right)
\label{nutplus2}
\end{eqnarray}

\section{Results and discussions}

\begin{figure}
\includegraphics[width=12cm,height=10cm]{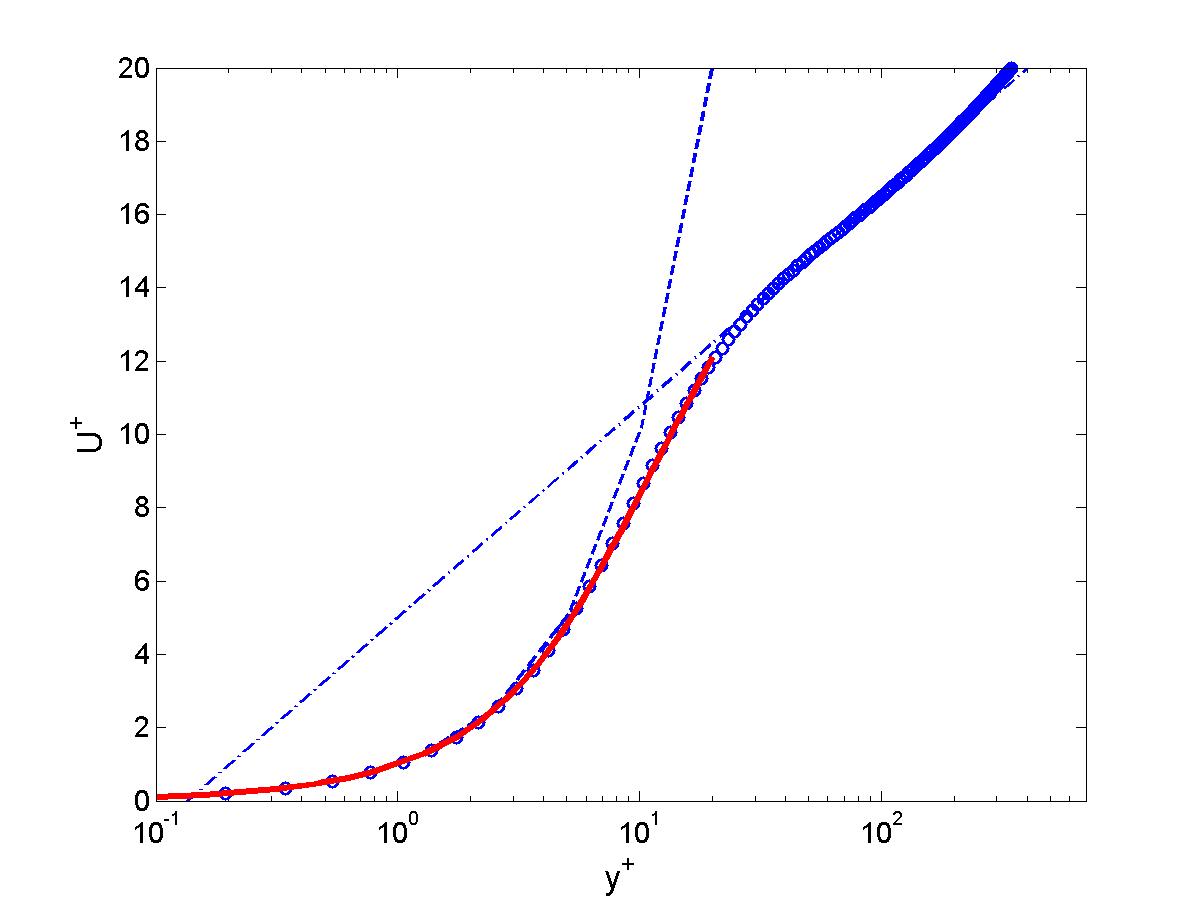}  
\caption{\label{fig:Usol} Mean streamwise velocity profile $U^+(y^+)$ for $Re_{\tau} = 642$. o, DNS data \cite{Iwamoto}. Curves: bold red solid line, solution of Eq. (\ref{NSF}) with Eq. (\ref{nutplus2}) ($C_{\nu}=0.3$, $A_l^+=26$, $A_k^+=8$ and $B=0.14$); dashed line, $U^+ = y^+$; dash-dotted line, $U^+ = 2.5 \:  ln(y^+)+5.0$. 
}
\end{figure}

\begin{figure}
(a) \includegraphics[width=7cm,height=7cm]{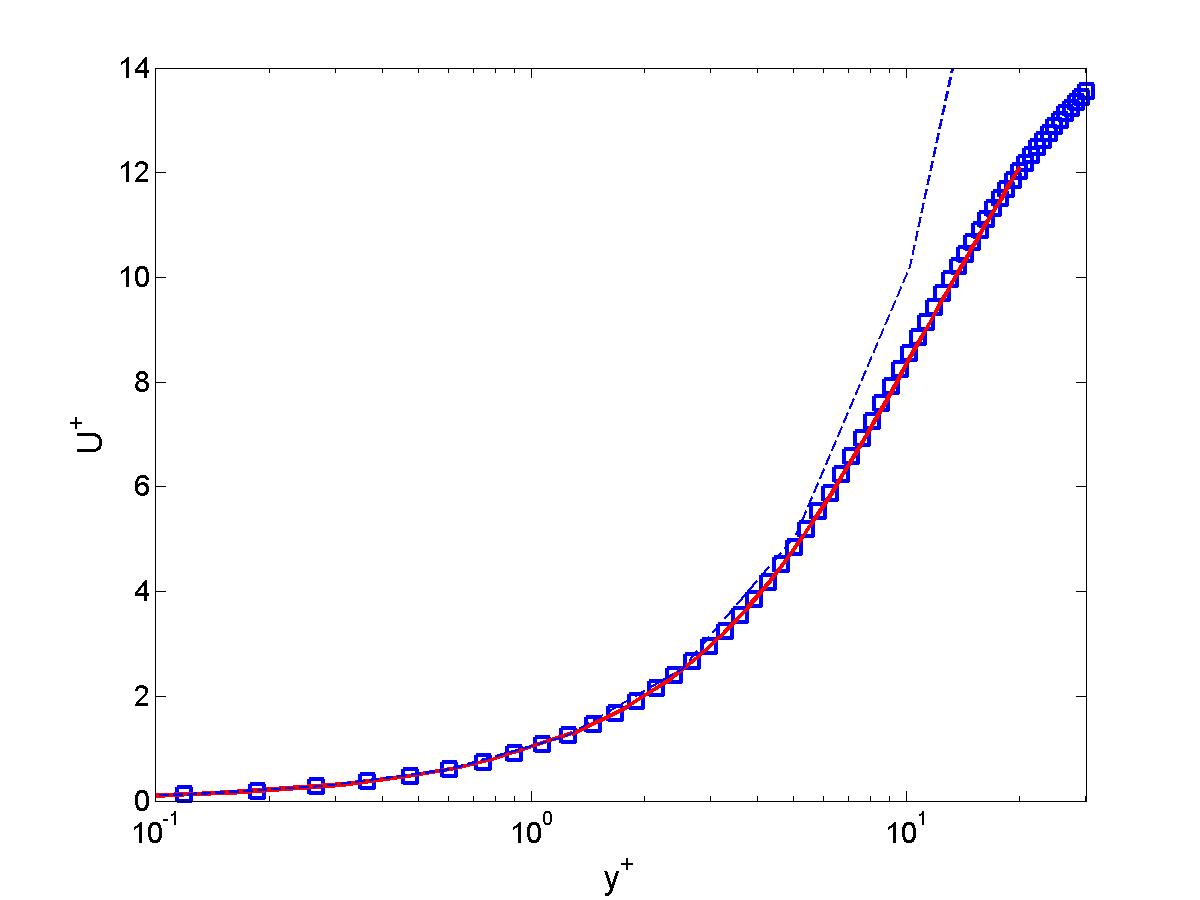} 
(b) \includegraphics[width=7cm,height=7cm]{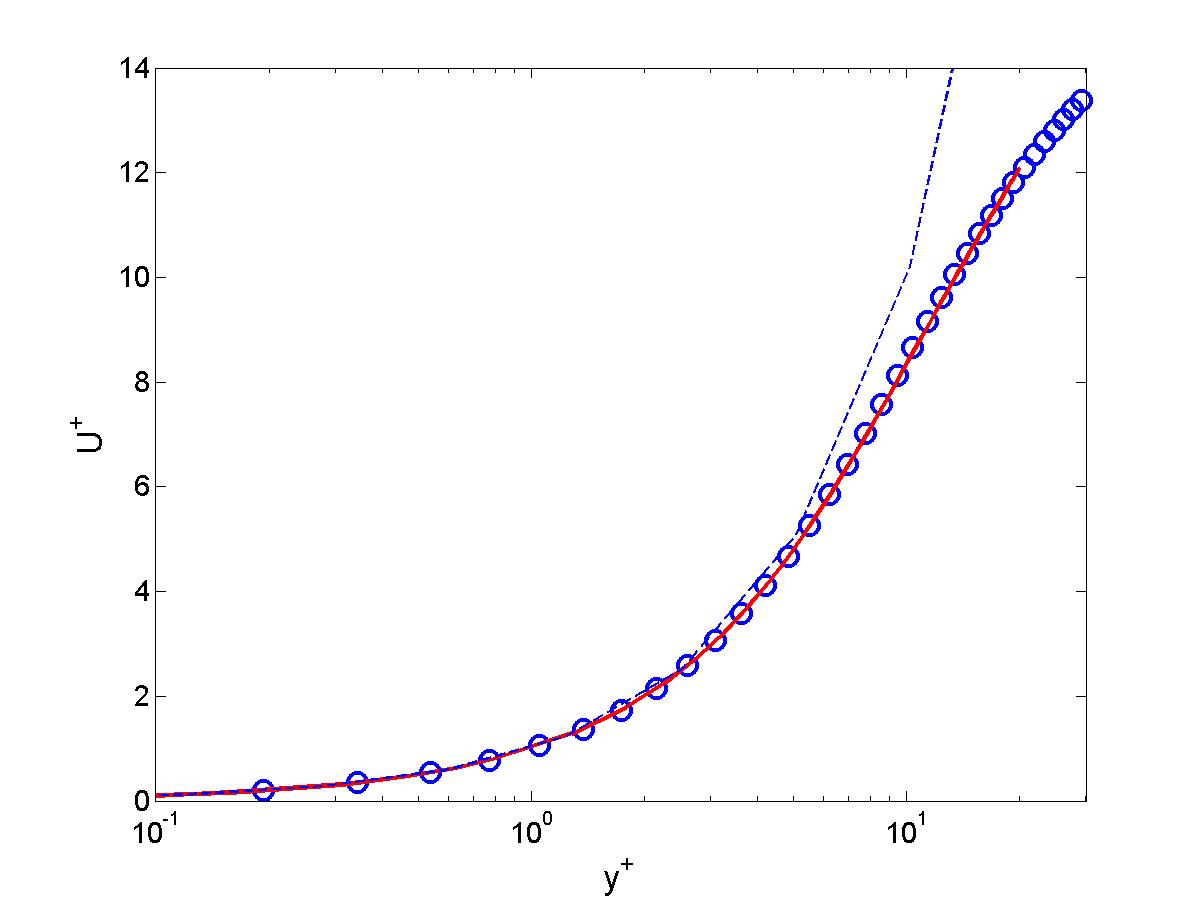}\\
(c) \includegraphics[width=7cm,height=7cm]{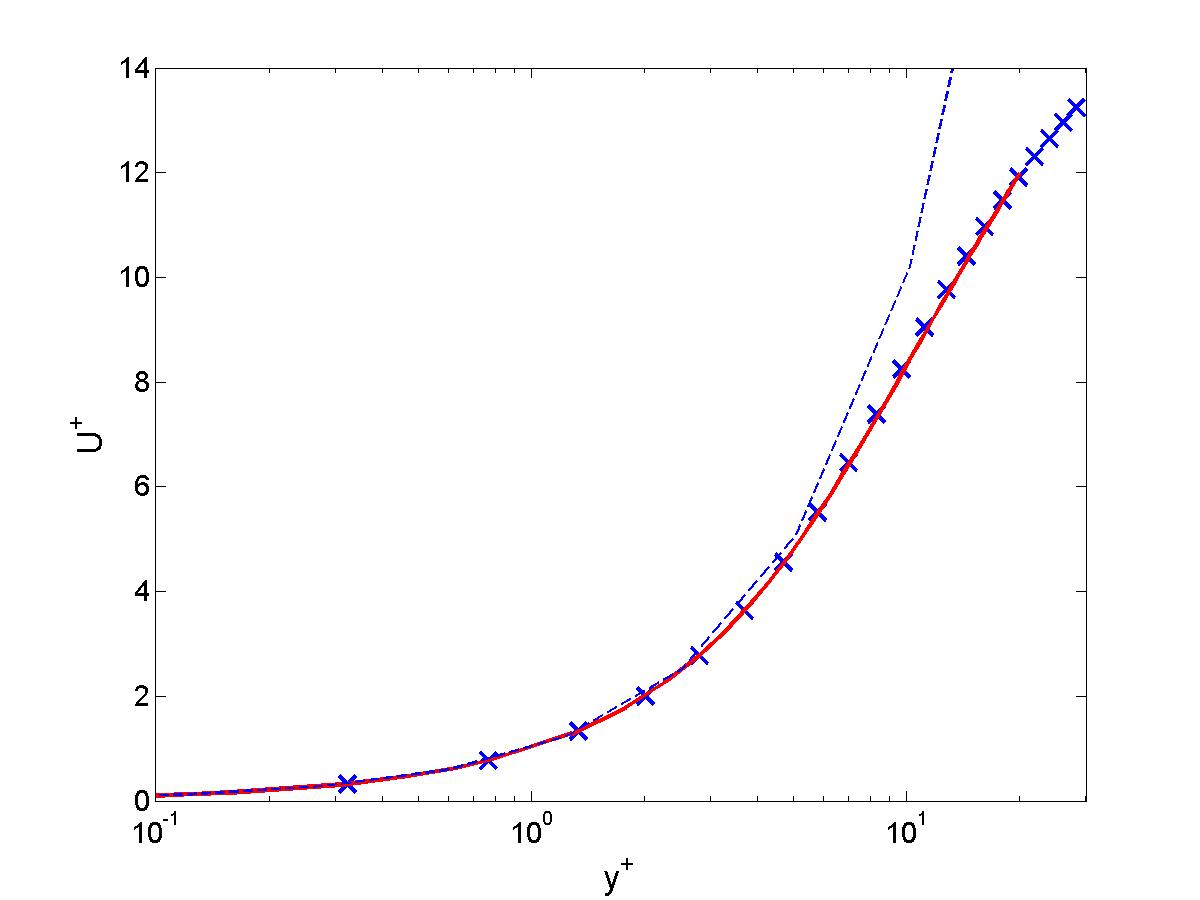}
\caption{\label{fig:Usol3} Mean streamwise velocity profiles $U^+(y^+)$  for $Re_{\tau} \geq 395$. 
Symbols, DNS data. Curves, Thin dashed line, $U^+ = y^+$. Red solid lines, solution of Eq. (\ref{NSF}) with Eq. (\ref{nutplus2}) ($A_l^+=26$, $A_k^+=8$); 
(a) $Re_{\tau} = 395$, squares \cite{Iwamoto}, solid line ($C_{\nu}=0.3$, $B=0.132$); 
(b) $Re_{\tau} = 642$, circles \cite{Iwamoto}, solid line ($C_{\nu}=0.3$, $B=0.14$); 
(c) $Re_{\tau} = 2003$, $\times$ \cite{Hoyas}, solid line ($C_{\nu}=0.3$, $B=0.158$). 
}
\end{figure}

\begin{figure}
(a) \includegraphics[width=7cm,height=7cm]{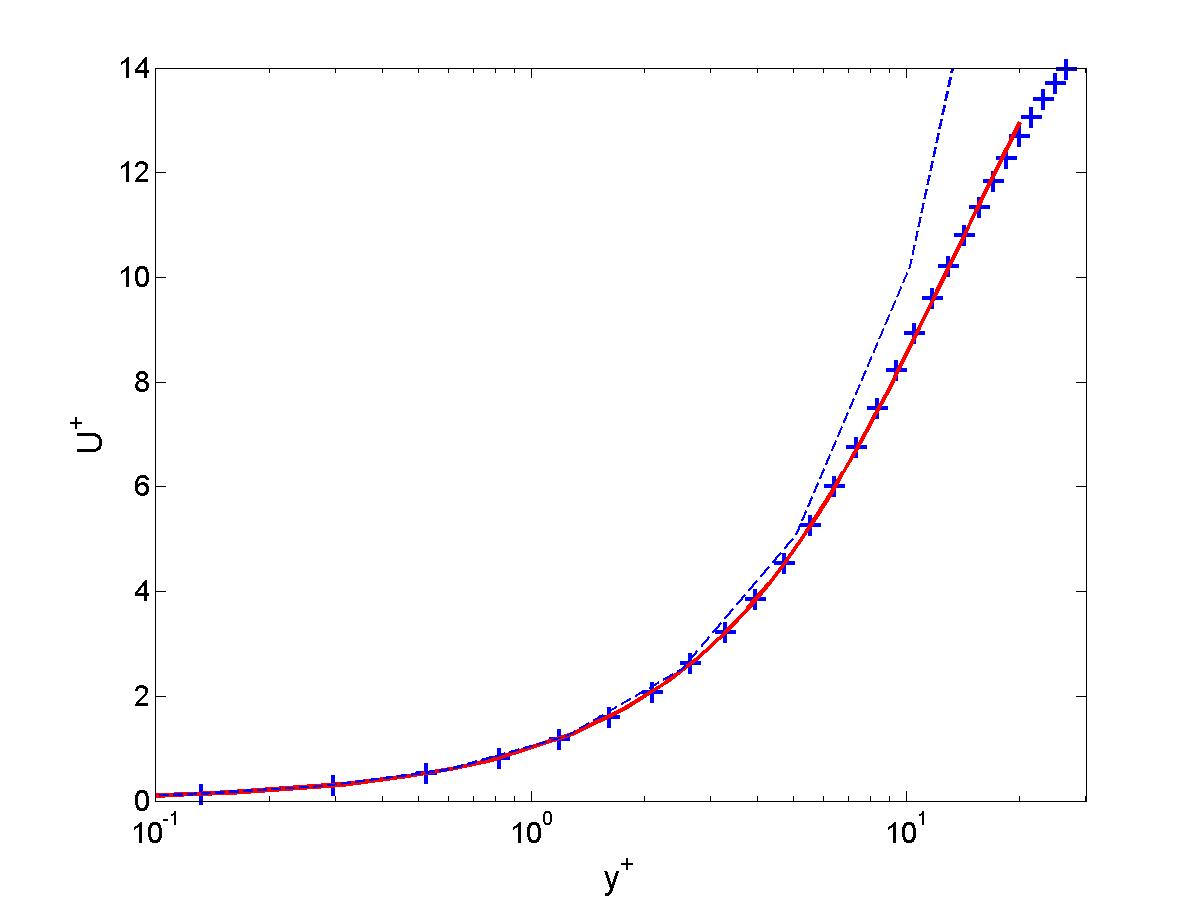}
(b) \includegraphics[width=7cm,height=7cm]{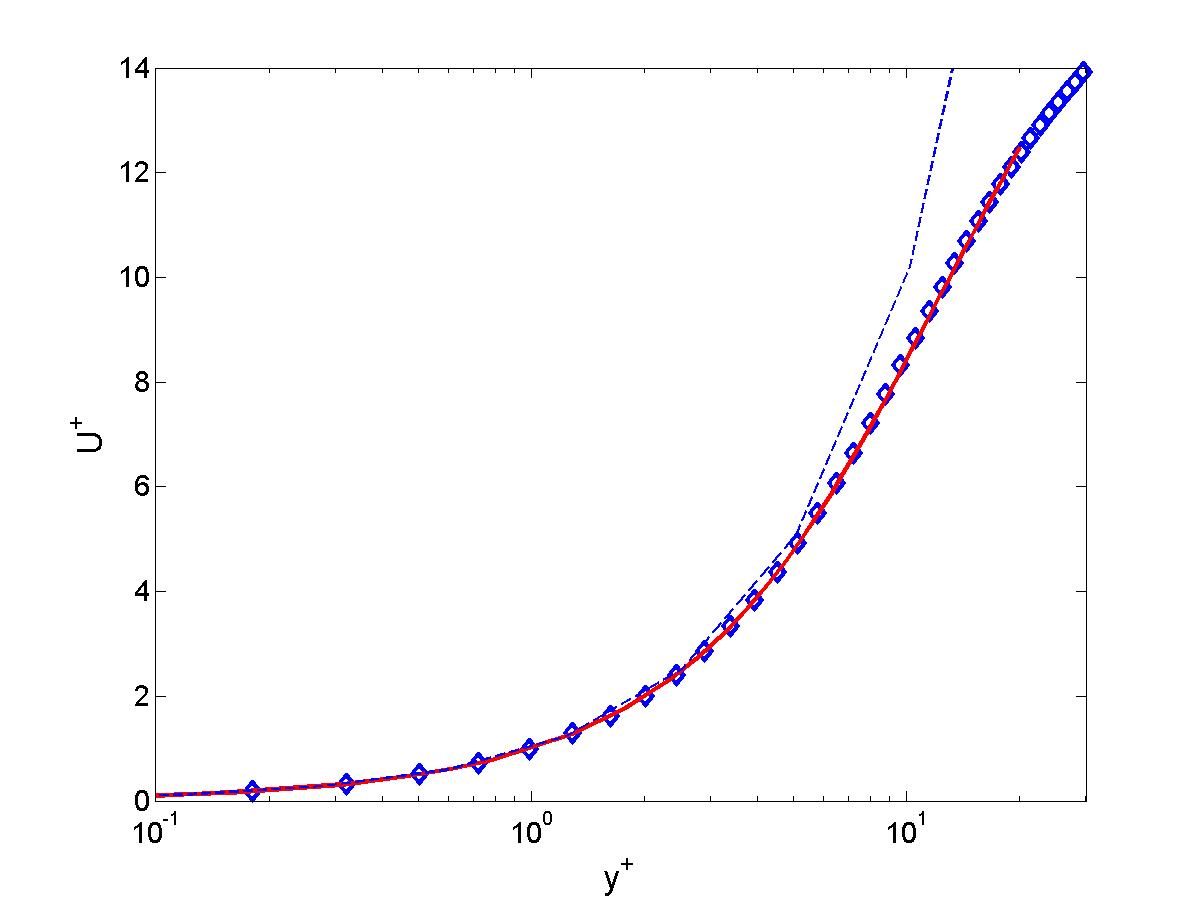}
\caption{\label{fig:Usol2} Mean streamwise velocity profiles $U^+(y^+)$  for $Re_{\tau} < 395$. 
Symbols, DNS data; Curves, Thin dashed line, $U^+ = y^+$; Red solid lines, solution of Eq. (\ref{NSF}) with Eq. (\ref{nutplus2}) ($A_l^+=26$, $A_k^+=8$); 
(a) $Re_{\tau} = 109$, $+$ \cite{Iwamoto}, solid line ($C_{\nu}=0.2$, $B=0.11$); 
(b) $Re_{\tau} = 150$, diamonds \cite{Iwamoto}, solid line ($C_{\nu}=0.25$, $B=0.116$); 
}
\end{figure}

Predicted mean streamwise velocity $U^+(y^+)$ profiles are obtained from Eq. (\ref{NSF}) and Eq. (\ref{nutplus2}). 
Figure (\ref{fig:Usol}) presents the mean streamwise velocity profile $U^+(y^+)$ for $Re_{\tau} = 642$. The solution of Eq. (\ref{NSF}) with Eq. (\ref{nutplus2}), where $A_l^+=26$,  $A_k^+=8$, $B=0.14$ and $C_{\nu}=0.3$, is compared to DNS data \cite{Iwamoto}. 
The predicted $U^+(y^+)$ profile shows good agreement with DNS data. 
Values of $A_k^+=8$ and $B=0.14$ are those of the $k^+$ profile (Fig. \ref{fig:kpyp}). 

In order to verify the dependency of the coefficient $C_{\nu}$ on the Reynolds number $Re_{\tau}$, we present predicted $U^+(y^+)$ profiles for different $Re_{\tau}$ (Fig~\ref{fig:Usol3}). 
Profiles of figure (\ref{fig:Usol3}) for $Re_{\tau} = 395$, $Re_{\tau} = 642$ and  $Re_{\tau} = 2003$ was obtained with $C_{\nu}=0.3$ and values of $A_k^+=8$ and $B$ from the $k^+$ profiles (Fig. \ref{fig:kpyp}). 
It seems that $C_{\nu}$ is independent of the Reynolds number for $Re_{\tau} \geq 395$ and is equal to $0.3$. The values of $B(Re_{\tau})$ obtained from the $k^+$ profiles are suitable for computation of $U^+(y^+)$ profiles. 
However, for $Re_{\tau} = 150$ and $Re_{\tau} = 109$ (Fig~\ref{fig:Usol2}) the required values of $C_{\nu}$ are respectively $0.25$ and $0.2$. Therefore, $C_{\nu}$ seems to be $Re_{\tau}$-dependent for $Re_{\tau}$ less than $395$. 
This dependency seems to be associated to low-Reynolds-number effects. Indeed, Moser \textit{et al.} \cite{Moser} showed that low-Reynolds-number effects are absent for $Re_{\tau} > 390$. 
We notice that for $y^+ < 20$, the required $C_{\nu}$ is different from  $C_{\mu}^{1/4}$ (with $C_{\mu}$ is the empirical constant in the $k$-$\epsilon$ model equal to $0.09$). For $Re_{\tau} \geq 395$, $C_{\nu} = C_{\mu}^{1/2}$.


\section{Conclusion}

In summary, 
mean streamwise velocity profiles $U^+$ was obtained by solving a momentum equation which is written as an ordinary differential equation. The analytical eddy viscosity formulation is based on an accurate near-wall function for the turbulent kinetic energy $k^+$ and the van Driest mixing length equation. The parameters obtained from the calibration of $k^+$ was used for the computation of $U^+$. 
Comparisons with DNS data of fully-developed turbulent channel flows show good agreement. 
Our simulations show that for $Re_{\tau} > 395$ the coefficient of proportionality $C_{\nu}$ in the eddy viscosity equation is independent of $Re_{\tau}$ and equal to $0.3$. However, for $Re_{\tau} < 395$, the coefficient $C_{\nu}$ is $Re_{\tau}$-dependent. 
The values of $k^+(y^+=20)$ and $U^+(y^+=20)$ 
could be used as boundary conditions 
for a turbulence closure model applied for $y^+ \geq 20$.\\ 

\textbf{AKNOWLEDGEMENTS} \\ 
The author would like to thank 
N. Kasagi, K. Iwamoto, Y. Suzuki from the University of Tokyo and 
J. Jimenez, 
S. Hoyas
from Universidad Polite´cnica de Madrid, 
for providing the DNS data. 

\end{document}